\documentclass[conference]{IEEEtran}
\makeatletter
\def\ps@headings{%
\def\@oddhead{\mbox{}\scriptsize\rightmark \hfil \thepage}%
\def\@evenhead{\scriptsize\thepage \hfil \leftmark\mbox{}}%
\def\@oddfoot{}%
\def\@evenfoot{}}
\makeatother
\usepackage{url}

\usepackage{breakurl} 
\usepackage[breaklinks]{hyperref}
\usepackage{graphicx}
\usepackage[usenames,dvipsnames,svgnames,table]{xcolor}
\usepackage{multirow}
\usepackage{cite}
\usepackage[ruled,vlined]{algorithm2e}
\usepackage{booktabs}
\usepackage{paralist}
\usepackage{subcaption}
\usepackage{mdwlist}
\usepackage{rotating}
\usepackage{mdframed}
\usepackage[font=small,skip=4pt,belowskip=-1pt]{caption}
\usepackage[all]{nowidow}
\usepackage{tikz-qtree}
\usetikzlibrary{matrix,shapes,arrows,positioning,chains,calc}
\usepackage[letterpaper]{geometry}
\renewcommand{\baselinestretch}{0.95}
\newcommand{\myparagraph}[1]{\noindent{\bf#1}}

\begin{document}

\title{Efficient Gossip Protocols for Verifying the Consistency of Certificate Logs}

\author{
\IEEEauthorblockN{Laurent Chuat}
\IEEEauthorblockA{ETH Zurich}
\and
\IEEEauthorblockN{Pawel Szalachowski}
\IEEEauthorblockA{ETH Zurich}
\and
\IEEEauthorblockN{Adrian Perrig}
\IEEEauthorblockA{ETH Zurich}
\and
\IEEEauthorblockN{Ben Laurie}
\IEEEauthorblockA{Google Inc.}
\and
\IEEEauthorblockN{Eran Messeri}
\IEEEauthorblockA{Google Inc.}
}

\maketitle

\begin{abstract}
The level of trust accorded to certification authorities has been decreasing
over the last few years as several cases of misbehavior and compromise have been
observed. Log-based approaches, such as Certificate Transparency, ensure that
fraudulent TLS certificates become publicly visible. However, a key
element that log-based approaches still lack is a way for clients to verify that
the log behaves in a consistent and honest manner. This task is challenging due to
privacy, efficiency, and deployability reasons.
In this paper, we propose the first (to the best of our knowledge)
gossip protocols that enable the detection of log inconsistencies.
We analyze these protocols and present the results of a simulation
based on real Internet traffic traces. We also give a deployment plan,
discuss technical issues, and present an implementation.
\end{abstract}

\section{Introduction}
\label{sec:intro}
A public-key infrastructure (PKI) and cryptographic protocols, such as the
widely deployed TLS, are crucial elements for many applications on today's
Internet, as they enable users to communicate sensitive data in
a---supposedly---secure manner. Unfortunately, since the inception of TLS,
many severe attacks have been
reported~\cite{clark:sok,comodo-attacks}. These attacks concern not
only the protocol itself, but also the oligarchical trust model that TLS relies on.
Certification authorities~(CAs) are usually considered as trusted by web
browsers when their self-signed certificates are present in the default list that
major vendors provide with their software.  These CAs are numerous (e.g., more than
one thousand CA public keys are trusted by Windows or
Firefox~\cite{eckersley2010observatory}). Attackers can exploit vulnerabilities and
potentially compromise private keys in order to obtain fake certificates, and
CAs can also intentionally produce certificates that will not be used by the
genuine owner of the corresponding domain. In particular, government agencies
that run or collaborate with CAs have the ability to impersonate certain
domains and intercept decrypted traffic through man-in-the-middle (MITM)
attacks~\cite{certified-lies}. For example, cases of unauthorized certificates
issued by trusted intermediary CAs have been reported in Turkey~\cite{turkey_mitm}
and France~\cite{france_mitm}.
To mitigate these problems, several approaches based on the concept of a
monitored public log of certificates have been proposed. They aim at making the
issuance process more transparent and accountable.
However, introducing a log as a new trusted third party moves the initial
problem rather than resolves it and raises the classical question
\textit{``Quis custodiet ipsos custodes?''~(Who watches the watchmen?)} An ideal solution
would give users a way
of verifying that they all share a consistent view of the log, on a worldwide
scale. Indeed, an attacker who can obtain a fraudulent certificate and
launch a MITM attack may also have the ability to control the log and provide a
view that contains a specific certificate only to targeted victims. 
This attack will be referred to as a \emph{split-world attack}~\cite{Mazieres:2002:BSF:571825.571840}.
To better illustrate it,
we can take the example of a malicious government that performs a MITM attack on
its citizens. To prevent this from happening, it is crucial that clients have a
means of exchanging information about the log, even if our hypothetical government
tries to prevent anyone from reporting the attack.

A \textit{gossip protocol} (i.e., a protocol in which
peers select partners in a somewhat random fashion in order to propagate
messages over the whole network) seems to be a promising way of solving
this problem, as it would be efficient and convenient to deploy.
However, we note that the term "gossip protocol" must not be taken strictly in
the traditional sense in this paper.  As the
protocol should not require a dedicated network or infrastructure,
peer selection is not completely random but client-driven, i.e., gossiping
is realized during standard HTTPS connections and servers are used as intermediaries
to exchange information from client to client.
The gossip protocol should guarantee that misbehavior is detected,
while privacy concerns are taken into consideration. The complex
structure of the Internet and the diverse behaviors of its users make developing
such a protocol a challenging task. Moreover, the gossip or epidemic-style protocols
that can be found in the literature~\cite{Voulgaris:2003:RSP:2155716.2155725, 1176982} are not
applicable in this context, as they generally require membership management
or an overlay network, instead of pre-existing connections.

\begin{inparaenum}[\itshape a\upshape)]
The main contributions of this paper are:~\item We present different
gossip protocols that meet the conditions stated above and we demonstrate
the properties that they exhibit.~\item We derive results about these
protocols from a simulation framework that we developed with the objective of
having an accurate model of Internet traffic, using both public and private sources of
statistical data that we had at our disposal.~\item We discuss deployment,
and implementation issues.~\item We identify the possible attack scenarios and
show how the presented protocols can detect them.~\item We describe a proof-of-concept
prototype of the system and show performance results.
\end{inparaenum}

\section{Background}
\label{sec:pre}
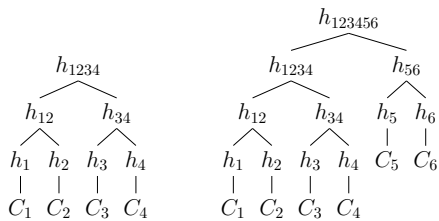
\begin{figure}
\begin{center}
\begin{tikzpicture}[scale=.6,every tree node/.style={font=\Large,anchor=base}]
    \Tree
     [.${h_{1234}}$
        [.${h_{12}}$
            [.$h_1$ $C_1$ ] [.$h_2$ $C_2$  ] ]
        [.${h_{34}}$
            [.$h_3$ $C_3$ ] [.$h_4$ $C_4$ ] ]
    ]
\end{tikzpicture}
\qquad
\begin{tikzpicture}[scale=.6,every tree node/.style={font=\Large,anchor=base}]
    \Tree
 [.$h_{123456}$
     [.${h_{1234}}$
        [.${h_{12}}$
            [.$h_1$ $C_1$ ] [.$h_2$ $C_2$  ] ]
        [.${h_{34}}$
            [.$h_3$ $C_3$ ] [.$h_4$ $C_4$ ] ]
    ]
        [.${h_{56}}$
            [.${h_5}$ $C_5$ ] [.$h_6$ $C_6$  ] ]
    ]
\end{tikzpicture}
\end{center}

\caption{Example of two versions of a ChronTree. Proof that $C_4$ is present in
the second tree (given the root $h_{123456}$): $\{h_3, h_{12}, h_{56}\}$.
Proof that the second tree is an extension of the first one (given roots
$h_{1234}$ and $h_{123456}$): $\{h_{56}\}$.}
\label{fig:trees}
\end{figure}

\myparagraph{Merkle hash trees.}
The important data structure used in many public-log systems is the
\textit{Merkle tree}, also called \textit{hash tree},
or just \textit{tree} throughout this paper.  More specifically, logs are
usually composed of a binary version of this tree in which leaf nodes are
hashes of some data (e.g., certificates) and non-leaf nodes are hashes of their
two concatenated children. Hash trees have interesting properties for
constructing a secure and efficient log~\cite{crosby2009efficient}. In
particular, knowing the \textit{root} (also called \emph{head}) of a tree, one
can prove that a leaf is part of it, with a number of nodes logarithmically
proportional to the number of leaves. Provided that the tree is
maintained in an append-only manner and that entries are added in chronological
order (sometimes called \textit{ChronTree}), the same holds for proving that a
hash tree is contained in (or contains) another hash tree. Figure~\ref{fig:trees}
shows an example.

\myparagraph{Certificate Transparency (CT)}\cite{rfc6962} 
was among the first log-based approaches to employ a hash tree (more precisely a ChronTree),
in order to build a publicly-verifiable database of certificates. In this framework,
anyone can submit a certificate to the log server that responds with a promise
to incorporate the certificate into the tree within a time period referred to as the
\emph{Maximum Merge Delay} (MMD). This promise takes the form of a so-called
\emph{Signed Certificate Timestamp} (SCT) that must be
delivered with the certificate by all TLS servers supporting CT. 
The head of the most recent hash tree can be retrieved from a log as
a \emph{Signed Tree Head} (STH), which also contains the tree size and a
timestamp.  As the name suggests, the entire object is cryptographically
signed by the log server.
To verify that the log server is behaving normally, two kinds of proofs can be queried.
When a client receives an SCT, it can verify that the corresponding
certificate is indeed present in the log by asking for an \emph{audit
proof}. In addition, the
append-only property of the log can be corroborated by a \emph{consistency
proof} between two different tree sizes. At present, the specification of CT does not
require that those proofs be signed; they should only contain the minimal number
of tree nodes needed to derive a tree head that coincides with a known STH.
Although signed proofs could, in some cases, constitute evidence that a log is
misbehaving and thus help achieving the goal of our gossip protocols, we shall assume,
in this paper, that logs behave exactly as described in the current documentation of CT.
As of today, Google is already running multiple pilot logs and Chrome
supports CT. It is required that Extended
Validation (EV) certificates issued after 1 February 2015 be present in
some log(s) and be delivered with one or more SCT(s). Ultimately, SCTs will be
required for all certificates~\cite{ct-ev}. The documentation of CT suggests
that gossip protocols might be a way of verifying the consistency of logs.
However, no protocol has been proposed to date.

\myparagraph{Roles and entities.}
The different participants in the system can assume the following roles:
 \textit{Clients}, usually web browsers, establish connections to
 \textit{servers} (identified by a domain name) via TLS. Note that, in some
    cases, the term ``client'' might be used to refer to clients of the log,
    but this is generally clear from the context.
 \textit{Certification Authorities}~ (CAs) are responsible for the issuance of
    certificates.
 \textit{Logs} allow anyone to submit certificates and make these
    certificates publicly available.
 \textit{Monitors} inspect every new entry added to certain logs to verify
    that they behave correctly and to detect illegitimate certificates.
    They may even store entire copies of logs to fulfill their objective.
 \textit{Auditors} verify that the log is behaving in a consistent manner
    with some partial information that they have or that they fetch from the
    log.
Different roles (e.g., client and auditor, CA and monitor) may be
assumed by a single entity. Typically, CAs may want to monitor some logs in
order to verify that certificates are not incorrectly issued on their behalf.

\section{Model}
\label{sec:model}
\myparagraph{Assumptions.}
To simplify technical descriptions, we generally assume that \textit{there is only one log},
but our work can be extended to multiple-log scenarios by running several instances of the
protocols in parallel. We also consider that \textit{CT is fully deployed} (i.e., within every TLS
Handshake the server sends an appropriate SCT) and there is some fraction
of servers and clients that use our protocol. It is assumed that
these \textit{clients and servers do not remove or modify gossip-related data/software}
and that \textit{the cryptographic primitives that we use are secure}.

\myparagraph{Adversary model.}
We consider that an adversary can operate a malicious log and provide evidence that a
certificate is in the hash tree only to specific clients (\textit{split-world attack}),
with the objective of hiding traces of this certificate to other clients of the log
that are potentially monitoring it. In order to degrade the efficiency
of the gossip protocol, the adversary can introduce malicious clients/servers and
inject chosen messages into the gossip protocol.
We also consider the simpler case where the view is consistent, but
an SCT is produced by the log and the corresponding certificate is never added
to the tree.

\section{Desired Properties}
\label{sec:properties}
The crucial property we seek is the \textit{detectability} of a
log's misbehavior, i.e., if the log presents a different set of certificates
to different clients at a given point in time, if it does not respect the
append-only property, or if, having produced an SCT, it fails to add the
corresponding certificate to the tree within one MMD, then the gossip protocol
should detect it. The speed of detection must be considered,
but is not as important as the guarantee that any malicious behavior will
eventually be detected with high probability. As the environment of the gossip
protocol is the Internet, the protocol should be \textit{scalable}, i.e., when
new clients or servers join the system, the solution should continue to reach its
goal as properly as before (if not better than before). The solution
should also be \textit{efficient}, in terms of \emph{storage}, \emph{computation},
and \emph{communication}  (the number of out-of-band connections should be
minimized, in particular, connections to the log server). Another important property
is \textit{deployability}. The protocol must not require an additional
infrastructure or an overlay network, and the deployment must be
done seamlessly via a regular software update.
Moreover, as the deployment of such a system would likely be incremental, it
must be determined how effective the solution is in function of the number of
clients and servers supporting the protocol.  Additionally, the protocol must
preserve users' \textit{privacy}.  In other words, it must not be possible to
determine which websites a user has visited based on the content of the
protocol's messages. 

\section{General Framework}
\label{sec:framework}
\begin{figure}[h!]
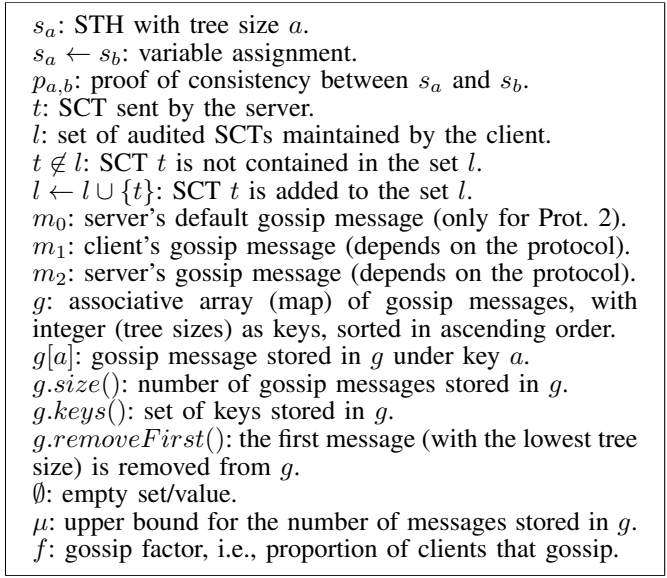


\begin{mdframed}
$s_a$: STH with tree size $a$. \\
$s_a \leftarrow s_b$: variable assignment. \\
$p_{a,b}$: proof of consistency between $s_a$ and $s_b$. \\
$t$: SCT sent by the server. \\
$l$: set of audited SCTs maintained by the client. \\
$t \not \in l$: SCT $t$ is not contained in the set $l$. \\
$l \leftarrow l \cup \{t\}$: SCT $t$ is added to the set $l$. \\
$m_0$: server's default gossip message (only for Prot. 2). \\
$m_1$: client's gossip message (depends on the protocol). \\
$m_2$: server's gossip message (depends on the protocol). \\
$g$: associative array (map) of gossip messages, with integer (tree sizes) as
keys, sorted in ascending order. \\
$g[a]$: gossip message stored in $g$ under key $a$. \\
$g.size()$: number of gossip messages stored in $g$. \\
$g.keys()$: set of keys stored in $g$. \\
$g.removeFirst()$: the first message (with the lowest tree size) is removed from
$g$. \\
$\emptyset$: empty set/value. \\
$\mu$: upper bound for the number of messages stored in $g$. \\
$f$: gossip factor, i.e., proportion of clients that gossip.
\end{mdframed}
\vspace{4pt}
\caption{Notation.}
\label{fig:notation}
\end{figure}

In this section, we give a high-level overview of message flows and describe the
actions performed by the different parties. Figure~\ref{fig:notation} presents
the notation that we will use.

All parties can query the log with the following functions:

\texttt{getSTH()}: returns the latest STH.

\texttt{getConsistencyProof($a, b$)}: returns a consistency proof between STHs
with tree sizes $a$ and $b$.

\texttt{getAuditProof($t, a$)}: returns the audit proof for the certificate
corresponding to an SCT $t$ and a tree size $a$.

The only way to hold perfect evidence that the log is misbehaving is to have two
(or more) STHs (with a valid signature) and observe that they are inconsistent.
To detect this, we will be using the following function:

\texttt{checkSTHs($s_a, s_b, ...$)}: STHs passed as arguments (or STHs contained
in more complex pieces of data, such as gossip messages, given as arguments)
meet the two following criteria: Two STHs with the same tree size have the same
root hash, and an STH with a larger timestamp than that of another STH has a
larger or equal tree size.

This function does not give any output. If the verifications passes, the
protocol can simply go on, but if a verification fails, the normal protocol flow
immediately stops and the inconsistency is reported/gossiped (as it will be
described in~\S\ref{sec:report}). Also, clients and servers always verify that a
received gossip message is valid with:

\texttt{validMessage($m_1$)}: returns a boolean value indicating whether $m_1$
is non-empty and valid (according to the message format definition of the given
protocol in \S\ref{sec:protocols}).

\begin{figure}[h!]
\centering

\begin{subfigure}[b]{\columnwidth}
    \centering
    \includegraphics[width=0.92\columnwidth]{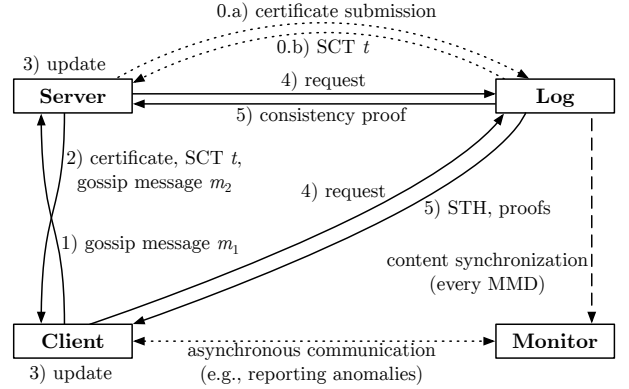}
    \caption{\footnotesize{Solid lines represent mandatory messages, dotted
    lines are optional flows, and the dashed line represent the monitor's role
    (see \S\ref{sec:pre} or~\cite{rfc6962}).}}
\end{subfigure}

\vspace{10pt}

\begin{subfigure}[b]{\columnwidth}
    \centering
    \includegraphics[width=0.92\columnwidth]{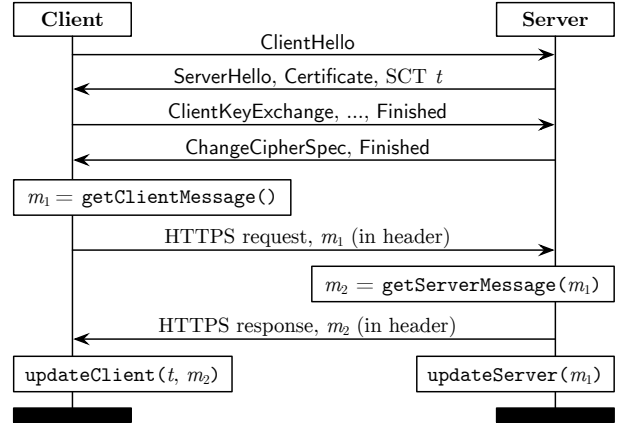}
    \caption{\footnotesize Detailed view of the client-server message exchange
    in steps 1--3.}
\end{subfigure}

\vspace{4pt}
\caption{Overview of communication flows.}
\label{fig:flows}
\end{figure}

A simple execution of the whole protocol is presented in Figure~\ref{fig:flows}.
The initial (one-time) operation conducted by a server's owner is to contact the
log, submit the certificate, and obtain an SCT in exchange. Afterwards, the
server should send this SCT during the establishment of all TLS connections.
Gossip messages are then \textit{piggybacked} on the requests and responses of
client-driven HTTPS traffic. Before a client connects to a server, a gossip
message $m_1$ is selected with the function $m_1 =$ \texttt{getClientMessage()}.
This message is sent along the client's HTTPS request. The server receives this
message and selects its own message accordingly with $m_2 =$
\texttt{getServerMessage($m_1$)}. Then, both parties update their local state:
\texttt{client\-Update($t, m_2$)} and \texttt{serverUpdate($m_1$)}. Note that,
for efficiency concerns, the response must be selected immediately and the
update procedure must be performed after the response is selected (and possibly
after it is sent). Depending on the received message, it might be necessary to
request a proof from the log to complete the update procedure. If there are
multiple logs in the system, the protocol can be executed in parallel for each
log, and several gossip messages (one per log) can be sent. If an anomaly is
detected, it must be reported to an appropriate entity, which we will assume to
be a monitor.

\begin{inparaenum}[\itshape a\upshape)] As the exchange is done through HTTPS,
the client receives (besides $m_2$) the server's certificate and a corresponding
SCT $t$. As gossiping SCTs may be inefficient and cause privacy issues (there is
a unique SCT for each certificate in the log), we choose to gossip STHs. Upon
receipt of an SCT, there are several pieces of data that clients must fetch from
the log \item the latest STH (that should be stored), \item a consistency proof
between the latest STH and a previous one (if available), and \item an audit
proof---for the certificate that corresponds to the received SCT---against the
tree size of the latest STH. The whole process can then either be repeated every
time an SCT is received or the SCTs (for which the operation has already been
accomplished) can be saved. These two possibilities yield opposite outcomes in
some aspects. The former allow clients to have a more recent view of the log.
The latter, instead, may leave clients with an outdated view of the log but
requires a much smaller number of connections. \end{inparaenum}

\section{Protocols}
\label{sec:protocols}
\subsection{STH-Only Gossiping}
\label{sec:sth_only}

In order to design an efficient and secure protocol,
one important observation needs to be made: any inconsistency that can be detected with
an STH of a certain tree size can also be detected with an STH of larger tree
size---provided that the consistency between these two STHs has been proven.
In other words, if a tree extends another tree, the smaller one can be
discarded without decreasing the likelihood of detecting an attack.
Based on this principle, we devise an STH-only gossip protocol, i.e., in which a valid
gossip message simply consists of a valid STH (a non-empty STH, in the right format, and
whose signature is verified with the known log's public key). 
As soon as clients obtain an STH from the log (as described
in~\S\ref{sec:framework}), they store it and send it in all their gossip messages.
Servers and clients only keep the STH with the largest tree size they encountered
and send it in all their gossip messages.
When clients receive an STH with a different tree size than that of the STH
they already have, they always contact the log for a consistency proof.
If the proof is correct and if the tree size of the received STH is larger than that
of the STH that is currently stored, then the client only keeps the received STH.
Clients keep track of SCTs and only contact the log when necessary (as discussed before).
Servers also communicate with the log to obtain consistency
proofs before discarding STHs. If all parties perform this verification, then a
strong alert signal can be produced when the log server is unresponsive.
This approach is detailed in Protocol~\ref{algo:protocolso}.

\begin{algorithm*}

\SetAlgorithmName{Protocol}{protocol}{}

\SetKwFunction{clientMessageSelection}{getClientMessage}
\SetKwFunction{clientUpdate}{clientUpdate}
\SetKwFunction{serverMessageSelection}{getServerMessage}
\SetKwFunction{serverUpdate}{serverUpdate}
\SetKwFunction{getConsistencyProof}{getConsistencyProof}
\SetKwFunction{getAuditProof}{getAuditProof}
\SetKwFunction{getSTH}{getSTH}
\SetKwFunction{check}{checkSTHs}
\SetKwFunction{validMessage}{validMessage}
\SetKwProg{func}{Function}{}{}
\SetKwProg{proc}{Procedure}{}{}

\centering
\fontsize{8.5}{8.5}\selectfont

\mbox{\parbox[t]{0.45\textwidth}{
    \centerline{\textbf{Client-side}}
    \BlankLine
    \textbf{Stored data}: STH $s_a$, set of audited SCTs $l$
    \BlankLine
    \func{\clientMessageSelection{} : \textnormal{message}}{
        \Return{$s_a$}\;
    }
    \BlankLine
    \proc{\clientUpdate{$t, m_2$}}{
    \KwIn{SCT $t$, server message $m_2$ $:= s_b$}
    \BlankLine
        \If{\validMessage{$m_2$}}{
            \check{$s_a, s_b$}\;
            \If{$a \ne b$}{
                \getConsistencyProof{$a, b$}\;
                \If{$a < b$}{
                    $a \leftarrow b$\;
                }
            }
        }
        \If{$t \not\in l$}{
            $s_c \leftarrow$ \textbf{\getSTH{}}\;
            \check{$s_a, s_c$}\;
            \getAuditProof{$t, c$}\;
            \If{$(s_a \ne \emptyset)$ and $(a < c)$}{
                \getConsistencyProof{$a, c$}\;
            }
            $s_a \leftarrow s_c$\;
            $l \leftarrow l \cup \{t\}$\;
        }
    }
}}
\hspace{15pt}
\mbox{\parbox[t]{0.45\textwidth}{
    \centerline{\textbf{Server-side}}
    \BlankLine
    \textbf{Stored data}: STH $s_b$
    \BlankLine
    \func{\serverMessageSelection{$m_1$} : \textnormal{message}}{
        \KwIn{client message $m_1$ $:=$ $s_a$}
        \BlankLine
        \uIf{not \validMessage{$m_1$}}{
            \Return{$\emptyset$}\;
        }
        \Else{
            \Return{$s_b$}\;
        }
    }
    \BlankLine
    \proc{\serverUpdate{$m_1$}}{
        \KwIn{client message $m_1$ $:=$ $s_a$}
        \BlankLine
        \If{not \validMessage{$m_1$}}{
            \Return{}\;
        }
        \If{$s_b \ne \emptyset$}{
            \check{$s_a, s_b$}\;
            \If{$a \ne b$}{
                \getConsistencyProof{$a, b$}\;
            }
        }
        \If{$(s_b = \emptyset)$ or $(b < a)$}{
            $s_b \leftarrow s_a$\;
        }
    }
}}
\caption{STH-only gossiping \emph{(notation and auxiliary functions are defined in \S\ref{sec:framework})}}
\label{algo:protocolso}
\end{algorithm*}

\subsection{STH-and-Consistency-Proof Gossiping}
\label{sec:sth_and_proof}

We now design a gossip protocol with the same security properties
as the protocol described above, but we aim
at reducing the number of connections to the log, by gossiping
consistency proofs together with STHs. This is important not only for efficiency reasons,
but also to minimize the ability of an adversary to infer who is gossiping and who is not.
As it will be discussed in more details in~\S\ref{sec:implementation}, if gossip messages are
conveyed through HTTPS headers, they are encrypted and thus it is harder for a passive attacker
to determine whether a given client is gossiping or not. However, if the adversary can observe
the communication between the client and the log, then it can guess, depending on the traffic,
if the client gossips. This possibility can be reduced if the type and number of requests sent
by the client to the log is essentially the same whether the client is only CT-enabled or
gossiping.
This second approach is presented in Protocol~\ref{algo:protocolsc}.

\begin{algorithm*}

\SetAlgorithmName{Protocol}{protocol}{}

\SetKwFunction{clientMessageSelection}{getClientMessage}
\SetKwFunction{clientUpdate}{clientUpdate}
\SetKwFunction{serverMessageSelection}{getServerMessage}
\SetKwFunction{serverUpdate}{serverUpdate}
\SetKwFunction{getConsistencyProof}{getConsistencyProof}
\SetKwFunction{getAuditProof}{getAuditProof}
\SetKwFunction{getSTH}{getSTH}
\SetKwFunction{check}{checkSTHs}
\SetKwFunction{validMessage}{validMessage}
\SetKwProg{func}{Function}{}{}
\SetKwProg{proc}{Procedure}{}{}

\centering
\fontsize{8.5}{8.5}\selectfont

\mbox{\parbox[t]{0.45\textwidth}{
    \centerline{\textbf{Client-side}}
    \BlankLine
    \textbf{Stored data}: STH $s_a$, STH $s_b$, consistency proof $p_{a,b}$, set of audited SCTs $l$
    \BlankLine
    \func{\clientMessageSelection{} : \textnormal{message}}{
        \eIf{$(s_a = \emptyset)$ or $(s_b = \emptyset)$ or $(p_{a, b} = \emptyset)$}{
            \Return{$\emptyset$}\;
        }{
            \Return{$(s_a, s_b, p_{a, b})$}\;
        }
    }
    \BlankLine
    \proc{\clientUpdate{$t$, $m_2$}}{
        \KwIn{SCT $t$, server message $m_2$ $:= (s_c, s_d, p_{c, d})$}
        \BlankLine
        \If{\validMessage{$m_2$}}{
            \check{$s_a, s_b, s_c, s_d$}\;
            \If{$(b \ne c)$ and $(b \ne d)$}{
                \textbf{\getConsistencyProof{$b, d$}}\;
            }
            \If{$b < d$}{
                $s_a \leftarrow s_c$\;
                $s_b \leftarrow s_d$\;
                $p_{a, b} \leftarrow p_{c, d}$\;
            }
        }
        \If{$t \not\in l$}{
            $s_e \leftarrow$ \textbf{\getSTH{}}\;
            \check{$s_b, s_e$}\;
            \getAuditProof{$t, e$}\;
            
            \eIf{$(s_b \ne \emptyset)$ and $(b < e)$}{
                $p_{b, e} \leftarrow$ \getConsistencyProof{$b, e$}\;
                $s_a \leftarrow s_b$\;
                $s_b \leftarrow s_e$\;
                $p_{a, b} \leftarrow p_{b, e}$\;
            }{
                $s_b \leftarrow s_e$\;
            }
            
            $l \leftarrow l \cup \{t\}$\;
        }
    }
}}
\hspace{15pt}
\mbox{\parbox[t]{0.45\textwidth}{
    \centerline{\textbf{Server-side}}
    \BlankLine
    \textbf{Stored data}: STH with the largest tree size $s_n$, map $g$, default message
    $m_0$ \\
    \textbf{Configuration parameter}: storage limit $\mu$
    \BlankLine
    \func{\serverMessageSelection{$m_1$} : \textnormal{message}}{
        \KwIn{client message $m_1$ $:=$ $(s_a, s_b, p_{a, b})$}
        \BlankLine
        \uIf{$($not \validMessage{$m_1$}$)$ or $(g$.size() $= 0)$}{
            \Return $\emptyset$\;
        }
        \uElseIf{$b \in g$.keys()}{
            \Return{$g[b]$}\;
        }
        \Else{
            \Return{$m_0$}\;
        }
    }
    \BlankLine
    \proc{\serverUpdate{$m_1$}}{
        \KwIn{client message $m_1$ $:=$ $(s_a, s_b, p_{a, b})$}
        \BlankLine
        \If{not \validMessage{$m_1$}}{
            \Return{}\;
        }
        \If{$s_n \ne \emptyset$}{
            \check{$s_a, s_b, s_n, g[a], g[b]$}\;
            \If{$(a \ne n \ne b)$ and $(a \not\in g.keys()$ or $b \not\in g.keys())$}{
                $p_{b, n} \leftarrow$ \getConsistencyProof{$b, n$}\;
                $g[b] \leftarrow (s_b, s_n, p_{b, n})$\;
            }
        }
        \uIf{$(s_n = \emptyset)$ or $(n < b)$}{
            $n \leftarrow b$\;
            $m_0 \leftarrow m_1$\;
            $g[a]\leftarrow m_1$\;
        }
        \ElseIf{$n = b$}{
            $g[a]\leftarrow m_1$\;
        }
        \While{$g.size() > \mu$}{
            $g.removeFirst()$\;
        }
    }
}}
\caption{STH-and-consistency-proof gossiping \emph{(notation and auxiliary functions are defined in \S\ref{sec:framework})}}
\label{algo:protocolsc}
\end{algorithm*}

\myparagraph{Message format.}
Messages have the same format for both clients and servers. Let $s_a$ and $s_b$
be STHs for hash trees of sizes $a$ and $b$ respectively, where $a < b$, and
let $p_{a, b}$ be a consistency proof between tree sizes $a$ and $b$, then a
valid message $m$ is a triplet of the form: $m = (s_a, s_b, p_{a, b})$.
Clients and servers always verify that a given message is in the right
format~(with \texttt{validMessage()}).
This format is desirable for the following reasons. Since consistency
proofs are not cryptographically signed, one cannot confirm that a proof genuinely comes
from a certain log, but if the two corresponding STHs are provided with it, at least
it can be verified that it is valid, preventing attackers from flooding servers with
invalid proofs. Furthermore, with such a format, if a message is valid but results
in an inconsistency (e.g., two STHs have the same tree size but not the same root
hash), it can only mean that the log is misbehaving. If an invalid proof was sent
alone or with only one STH, it could mean, in some cases, that either the server or
the log is the origin of the problem, without any way for clients to determine which
one it really is.

\myparagraph{Storage.}
Clients store three elements that correspond to the content of a message,
namely two STHs $s_a$ and $s_b$ and a consistency proof $p_{a, b}$. However,
a valid message cannot be immediately constituted, because this data is
not available at the very first execution of the protocol.
Clients also keep a set of audited SCTs.
Servers keep a collection
of valid messages sent by clients. More
precisely, servers store messages in a map $g$.\footnote{To reduce the size of
this map and avoid duplicates, servers can store STHs and consistency proofs
separately and keep only references to these instead of entire messages.}
Servers also save the STH with the
largest tree size they encountered, denoted $s_n$
(initially set to $\emptyset$).

\myparagraph{Message selection.}
Selecting a message on the client-side simply consists of grouping the two STHs and
the proof if they are available, or returning an empty message
otherwise.
On the server-side, the way in which messages are stored allows a response to be
selected efficiently. Servers simply select the message in their map by using
the tree size of the STHs in the client's message as a key,
provided that such a message exists. Otherwise, a default
message $m_0$ is sent.

\myparagraph{Update procedure.}
If the server's message contains a proof that
allows the client to obtain an STH with a larger tree size, then the content of the
server's message replaces the client's previous data. Otherwise, a consistency
proof is needed. If the SCT delivered with the server's certificate has never been
verified by the client, then the log is contacted for an audit proof.
If the message received by the server is not valid, it is simply dropped.
Otherwise, the server keeps the message in its map under a key that corresponds
to the tree size of the first STH. If a similar message already exists, it is
replaced. When necessary, servers also request consistency proofs, but there are
two situations in which these proofs are not needed:
when one of the STHs in the received message correspond to the STH with
the largest tree size ($s_n$) and the consistency proof is valid, and
when both STHs in the received message are already stored (then the root hashes
can simply be compared).
If a message $m_1$ received by the server contains an STH with a larger
tree size than $n$, then $n$ is updated and the default message is set to $m_1$.
The message is stored if the tree size of the second STH in the message is equal to
or greater than $n$. An upper bound $\mu$ on the storage size must be defined as a
configuration parameter. When this limit is reached and a message is added,
the first entry of the map (i.e., the message referenced by the smallest
tree size) must be removed.

\section{Reporting/Gossiping Anomalies}
\label{sec:report}
We made the assumption that a monitor has the ability to receive and process
the anomalies reported by clients or servers. A monitor is the entity that fits best into this role,
since its function is to constantly verify that the log server is not misbehaving, by contacting it at
regular intervals and possibly by keeping an entire copy of the tree. However, it might not be
possible to reach this monitor at all times, either because it is blocked by an attacker or because of
a technical issue. This is why clients and servers should both report and gossip about anomalies as
soon as they are detected and as long as the problem is not fixed (e.g., until a proof becomes
available or until the log becomes untrusted).
Anomalies of different kinds can be detected in several circumstances. When it happens, the normal
execution of the gossip protocol must stop and a special message must be reported and sent in
place of all normal gossip messages. We define two special messages.

A \emph{warning message} (containing a description of the problem and, if applicable, the incriminated STH/proof) is
generated when:
\begin{inparaenum}[\itshape a\upshape)]
    \item The log is unresponsive (either because it deliberately does not produce any response
    or because of network issues), and it remains unresponsive after a number of attempts (specified as parameter) to
    contact it again.
    \item The latest STH has a timestamp older than one MMD before the time at which it was fetched.
    \item An STH/SCT with an invalid signature or an invalid proof is received from the log.
\end{inparaenum}

An \emph{inconsistency message} is generated when the \texttt{checkSTHs} function (as described above) fails. This message 
consists of two or more valid STHs (signed by the log) that mutually present an inconsistency concerning their tree size, root hash,
or timestamp.

The recipient of a warning message must independently verify, by contacting the log, that the described problem exists
(provided that such an observation, e.g. "the log is unresponsive", was not already made recently).
If the problem is confirmed, then it must be reported to the monitor---we may assume that the attacker cannot block this access
to all clients---and the message must be propagated further through gossiping.
On the other hand, when a valid inconsistency message is received, the log can immediately be considered as
untrustworthy and the message can directly be reported and propagated, as it contains sufficient cryptographic evidence. Also, an
inconsistency message will take priority over a warning message.

\section{Simulation}
\label{sec:simulation}
\subsection{Simulation Framework}\label{sec:simulation_framework}

The main challenge in evaluating a scheme such as the one we propose is to create
a reliable client-server connection model. One important obstacle is the scarcity of
authentic traces of real-world connections (as this data is
privacy-sensitive). However, the framework that we developed and that we describe in this section
emulates realistic connections on a worldwide scale. To be precise, the traffic
of 112 countries (for which sufficient statistical data was available) is
simulated. Only a fraction of clients execute the given gossip protocol. This
proportion of clients is referred to as the \emph{gossip factor}, denoted $f$.
The MMD is set to 2 hours, as for the first pilot logs run by Google and
exactly one new STH is produced by the log every MMD. In each country, the number
of Internet users was estimated with the total population~\cite{population2014} and
the percentage of individuals using the Internet~\cite{internet-usage2012}.

In order to determine how many connections users perform during the day and how
these connections are distributed among users, we used a private 24-hour-long trace of
real HTTP/HTTPS traffic from 2014 provided directly by \emph{SWITCH}
(major manager of networks among universities and research facilities in Switzerland).
This data contains more than 104 million entries for HTTP and more than 74 million
entries for HTTPS, where each entry is a triplet of the form: relative time
(in seconds), client ID (anonymized), and server ID (an\-onym\-ized).
For each hour, we approximated the parameters of a negative
binomial distribution\footnote{Another simpler model that can express the number
of events occurring in a period of time is the Poisson distribution. It relies
on a single parameter that corresponds to both the variance and the mean, but
this is not sufficient in our context since the variance
exceeds the mean to a large extent. The negative binomial is a
generalization of the Poisson distribution~\cite{cameron2013regression} that
allows the variance to be different from the mean.}
by using maximum-likelihood estimation.
Random numbers of connections can then be generated and used in the simulation.
As different types of traffic are generated during different periods
of the day (for each country), we also take time zones into consideration.

Then, it must be determined to which websites these users connect. Amazon's Alexa
Web Information Service provides a vast quantity of precious information in this
regard. In particular, we collected data (in June 2014) about the top 100 domains for each of
the 112 countries. This includes the number of page views per million, i.e., the
number of page views that a particular site generates among one million pages
that are viewed by typical users. Based on this distribution, a random domain can be
picked. We have also taken into consideration the
possibility that a client connects to a domain outside of the
top 100 by reducing the total number of connections that a client perform
proportionally.

We scanned the servers of all these top domains to determine not only if they support
HTTPS (e.g., by verifying that the port 443 is open), but also if their certificate is valid,
if a connection can be established, and if they automatically redirect clients
to HTTPS. Globally, this condition is met by 287 out of 5107 servers scanned
(i.e., around 5.62\%). In our simulation, all these servers (and only these) are
both CT-enabled and gossiping, unless stated otherwise. This assumption is both realistic, because
major websites (that already enforce the usage of TLS) are more likely to adopt new security
mechanisms rapidly; and sufficient, because these sites generate a substantial
part of the global Internet traffic (about 33.5\% of the page views of a country
on average).
The validity period of all certificates is set to 24 months (moderately below
the maximum of 39 months imposed by the CA/B Forum~\cite{CABForum} for certificates
issued after 1 April 2015). For each certificate, the date of issuance is chosen
uniformly at random in the 24 months preceding the start of the simulation.

\subsection{Results}\label{sec:simulation_results}

We show simulation results in four distinct cases. The first basic situation
in which no gossip protocol is used can be subdivided into two cases: when clients keep
track of the SCTs they received and verified, and when they do not. We 
compare these cases with the two protocols presented in \S\ref{sec:protocols}. All
results were collected during a simulation of 365 days, and the
gossip factor was set to 0.1\% (2,462,216 gossiping clients), unless stated otherwise.
Results are usually rounded to the nearest hundredth, except when more precision is needed.
The storage limit $\mu$ was set to 10,000 messages, but, as the storage requirements of
our protocols are low, this limit was never reached in the 365 days period.

\begin{figure}
\centering
\includegraphics[width=\columnwidth]{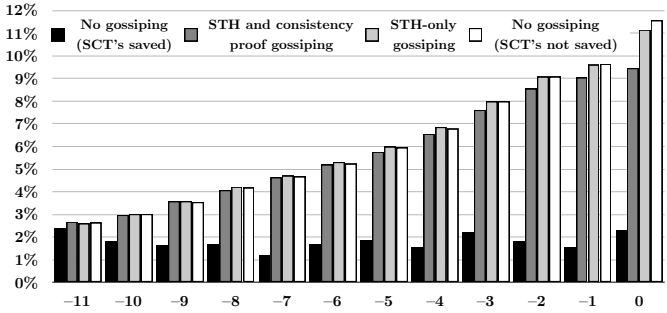}
\caption{Average distribution of the last 12 STHs among clients at the end of an
MMD.  On the horizontal axis, $0$ represents the last STH generated by the log,
$-1$ represents the STH that was generated 1 MMD before that, and so on.}
\label{fig:sth_distribution}
\end{figure}

\myparagraph{STH distribution.}
Since one of our strategies consists of keeping STHs with the
largest tree size, it is worth investigating how effective our protocols are in spreading
recent STHs. Figure~\ref{fig:sth_distribution} illustrates how the last 12 STHs (i.e.,
those that have been generated in the last 24 hours, since the MMD is set to 2) are distributed
as a percentage of clients having them stored at the end of an MMD.
We can observe that the two non-gossiping cases form a lower
and an upper bound for the distribution of the latest STH. Indeed, when SCTs
are saved, after some time, clients do not need to contact the log very often as they tend
to visit the same domains most of the time. The distribution is almost
uniform, because clients can keep the same STH as long as they do not connect to a website
they had never visited before. In this case, more than 78\% of the clients hold an STH 
that is more than one-day old. On the other hand, when SCTs are not saved, the same
information (including the latest STH) must be fetched from the log every time an HTTPS
connection is established with a CT-enabled server. Under these circumstances, 
only less than 26\% of gossiping clients do not hold one of the last 12 STHs.
This performance cannot be exceeded by a gossip protocol that relies on existing
client-driven connections.
There might be a bias in the results coming from the fact that we used a 24-hour trace of
HTTP(S) traffic. This trace does not necessarily contain some clients that establish connections
less often (e.g. once a month). The lack of longer authentic traces prevents us from modelling
the system more accurately; nevertheless, the presented results are relevant and undoubtedly
helpful to analyze the protocols.

Although distribution measurements are not directly useful to demonstrate how well
a protocol performs in detecting attacks, they can show how much clients collaborate to obtain
their vision of the log when combined with another metric. Indeed, there are
only two ways for clients to get a recent STH: contacting the log or receiving an appropriate
gossip message. If the distribution is close to the upper bound and the number of STHs
fetched from the log is
low, it means that the protocol achieves an efficient communication of log-related information
between clients, and attacks are more likely to be detected rapidly. The number of STH
queries~(\texttt{getSTH()} function) per MMD was measured to be about 7.02 million when SCTs
are not saved (one query for each HTTPS connection), and only about 10,000 when SCTs are saved
(this concerns both gossip protocols).

\myparagraph{Overhead.}
Another decisive characteristic that must be analyzed is the overhead that
gossip protocols introduce.
More precisely, we express the overhead as the number of log connections
strictly generated by the gossip protocol (not by the standard CT framework, i.e., we do not
consider queries for the latest STH and the audit proof, but only queries for consistency
proofs that are needed upon receipt of certain gossip messages) over the total number of
HTTPS connections performed by gossiping clients.
The average number of HTTPS connections per MMD was measured to be approximately 7.02 million
in all cases, and the overhead was about 8.58\% (602,263.2 log connections per MMD)
for the STH-only protocol and only about 0.0058\% (407.58 log connections) for
the second protocol, when proofs are gossiped. This constitutes a substantial overhead  reduction and shows, therefore, that gossiping consistency proofs is indeed pertinent.
In the first protocol, exactly half of the gossip-related queries to the log originate
from clients and the other half from servers, since they need to request consistency proofs
under the same conditions, i.e., when the STHs they exchange have a different tree size.
In contrast, when the second protocol is used, more queries (72.09\%) come from clients.

\myparagraph{Storage.}
For both gossip protocols, the storage required on the client-side (strictly for
gossiping) is fixed: a few kilobytes for, at most, a consistency proof and the pair of
corresponding STHs. Moreover, the average number of SCTs stored by clients was measured to be
about 11--12 (recall that only one hundred websites are considered for each country).
Considering that SCTs are only a few hundred bytes, this storage is almost negligible compared
to other data that browsers usually store: history, cookies, cached media files, and so on.
On the server-side, the second protocol needs to store gossip messages. The size of the map used
to store these messages is monotonically increasing and the maximum number of messages that
can be stored at a certain point in time is related to the number of different STHs that have
been generated by the log. We observed that the largest map contained 4380 messages at the end
of the 365 days of simulation. As an example, a limit of 10,000 messages would require less than
16 MB of storage while being able to hold more than two years worth of log data.

\begin{figure}
\centering
\includegraphics[width=\columnwidth]{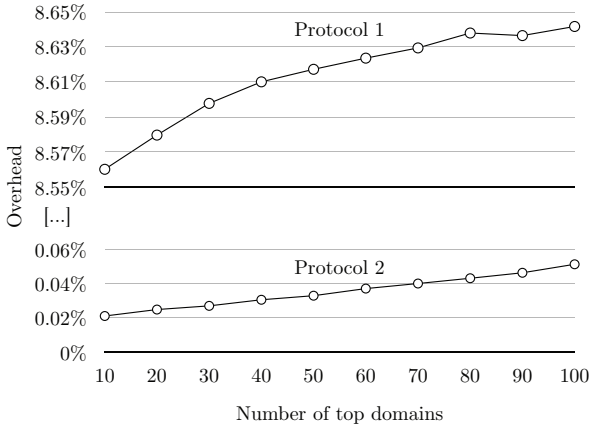}
\caption{Overhead in function of the number of gossiping HTTPS servers in the top $X$ domains of each country (the gossip factor is fixed to 0.01\%), for both gossip protocols.}
\label{fig:scalability}
\end{figure}

\myparagraph{Scalability.}
Table~\ref{tab:scalability} shows how the protocols
behave when we vary the gossip factor $f$. We also compare the situation in which HTTPS servers
in the top 100 domains are gossiping (as before) with the situation in which only the global top domain (google.com) is gossiping while HTTPS servers are still CT-enabled. We observe that both protocols perform better, in terms of distribution of recent STHs, when more clients or more servers participate and that the overhead of the second protocol gets even smaller when the gossip factor is increased. It is worth noting that even when only the global top domain (google.com) is gossiping, the protocol achieves a performance close to the results we presented before (with many more servers). This is explained by the fact that google.com is responsible for an important part of the total number of page views in all countries. Moreover, we remark that, although the distribution results are very similar
for both protocols, the overhead is substantially lower when proofs are gossiped.
Figure~\ref{fig:scalability} shows how the overhead evolves when we change the number
of gossiping servers. The second protocol not only generates a much lower overhead
than the first one, but we see that it scales well, as the overhead increases
steadily but slowly when more servers participate to the protocol.

\begin{table*}
\centering
\small
\renewcommand{\arraystretch}{0.95}

\begin{tabular}{@{}rm{60pt}rrrcrrr@{}}
\toprule
&\multirow{2}{*}{\vspace{-0.35em}Gossiping} & \multicolumn{3}{c}{Global top domain} & \phantom{abc} & \multicolumn{3}{c}{Top 100 domains for each country} \\
\cmidrule{3-5} \cmidrule{7-9}
&& $f=0.001 \%$ & $f=0.01 \%$ & $f=0.1 \%$ && $f=0.001 \%$ & $f=0.01 \%$ & $f=0.1 \%$ \\
\cmidrule{2-9}
\multirow{3}{*}{\begin{turn}{90}Prot. 1\end{turn}} & Latest STH & 6.06 \% & 8.9 \% & 9.03 \% && 6.07 \% & 9.86 \% & 11.13 \% \\
& Last 12 STHs & 47.67 \% & 66.88 \% & 66.79 \% && 54.18 \% & 72.1 \% & 73.98 \% \\
& Overhead & 6.71 \% & 6.77 \% & 6.77 \% && 9.09 \% & 8.64 \% & 8.58 \% \\
\cmidrule{2-9}
\multirow{3}{*}{\begin{turn}{90}Prot. 2\end{turn}} & Latest STH & 5.75 \% & 7.73 \% & 8.6 \% && 6.88 \% & 8.8 \% & 9.45 \% \\
& Last 12 STHs & 51.87 \% & 63.23 \% & 64.48 \% && 53.83 \% & 69.01 \% & 70.01 \% \\
& Overhead & $2.25 \cdot 10^{-3}$ \% & $2.2 \cdot 10^{-4}$ \% & $1.2 \cdot 10^{-5}$ \% && $3.8 \cdot 10^{-1}$ \% & $5.12 \cdot 10^{-2}$ \% & $5.81 \cdot 10^{-3}$ \% \\
\bottomrule
\end{tabular}

\smallskip{}
\caption{Average number of gossiping clients with the latest STH or any of the last 12 STHs, average overhead (as defined before), and storage usage defined as the average number of messages stored by servers over the number of different STHs generated by the log, for both gossip protocols. Different gossip factors $f$ and different sets of gossiping servers are considered.}
\label{tab:scalability}
\end{table*}

\section{Analysis and Discussion}
\label{sec:analysis}
There are two opposite scenarios of split-world attacks in which the
detectability property of our protocols can be analyzed. The first scenario is
the one in which a MITM attack targets a single client, using an illegitimate
certificate, and the log is controlled by the attacker (e.g., a government) to
provide this certificate only to the victim. If the client is gossiping and
connects to a non-compromised gossiping server, then there are two possibilities
regarding the tree sizes of the exchanged STHs.  If they are equal, then the
root hashes will not be the same and the attack will be detected. Those two STHs
are a strong evidence that the log is misbehaving, and anyone who receives them
can acknowledge the attack.  If the tree sizes are different, then both the
client and the server will request a consistency proof that the log will not be
able to produce.  This means that the log will either respond with an incorrect
proof or be unresponsive.  In both cases, an alert message will be gossiped and
possibly reported to some entity~(e.g., a monitor). Here, there is no tangible
evidence that the log is malicious, so all participants in the gossip protocol
should independently try to confirm the assertion contained in an alert message
they receive to produce a signal as strong as possible. The second attack
scenario is the one in which the whole population is divided into different
parts, for example, when a country performs a large-scale permanent MITM attack.
As soon as a gossiping client connects to a gossiping server that is not hit by
the MITM, the attack will be detected as before. Even with a country-wide
firewall, this is very likely to happen if at least one user travels to another
country with his device and connects to a server that supports the protocol. Of
course, this statement holds only if the number of gossiping clients and servers
is sufficient.

The speed of adoption for a new security feature such as the one we described
depends on the software update mechanism. As modern brows\-ers are usually
automatically updated at regular intervals, a fast deployment rate could be
guaranteed for clients. On the other hand, servers usually need to be updated
manually and the adoption of a new technique could be much slower. In any case,
this is not to be considered as a major issue as long as the protocol is
scalable and the price to pay in terms of additional connections and storage is
not too high, which was shown to be the case in \S\ref{sec:simulation_results}.

In order to reduce the overhead introduced by the gossip protocol even more,
consistency proofs should be requested in batch at regular intervals, instead of
immediately when needed. The same goes for verifying the claim contained in a
warning message. This will not only decrease the number of requests to the log,
but also avoid to fetch the same information several times in the defined period
of time. Attacks would only be detected slightly more slowly, depending on the
value at which the interval is set.

The gossiped pieces of data are usually not privacy-sensitive, as they should be
identical for all users and thus should not allow to infer which domains a
client visited. However, if the log fails to prove that a certificate was added
to the tree, the corresponding SCT must be reported and this causes privacy
issues. The SCT (tagged as, e.g., ``not present in the log'' or ``log
unresponsive'') could be sent back only to the server it came from, but this is
ineffective if the server is malicious, so it should be required that users give
their consent to gossip the fraudulent SCT.

\section{Realization in Practice}
\label{sec:implementation}
\myparagraph{Protocol layer (TLS vs. HTTPS).} To satisfy the deployability
requirement, we used HTTPS traffic to transport gossip messages. Hence, our
protocols can be implemented either in the TLS layer (as suggested
in~\cite{Laurie:2014:CT:2668152.2668154}) or in the HTTP layer. The first option
is to use a \textit{TLS extension} and to introduce a dedicated field in
\textit{ClientHello} and \textit{ServerHello} for the gossip messages. The
advantage of such a solution is that all TLS-supported services (e.g., SMTP over
TLS) can be used for gossiping. Unfortunately, this approach has some drawbacks
too.  First, ClientHello and ServerHello messages are sent unencrypted.
Therefore, an eavesdropper could determine whether a client is gossiping and
read the exchanged messages. In such a setting, an adversary could launch a MITM
attacks only on non-gossiping users. Moreover, too many additional bytes
introduced to ClientHello or ServerHello messages may cause latency in the TLS
Handshake~\cite{overclocking_ssl}. This last point is crucial in a multi-log
setting. The alternative approach is to use HTTP headers. The gossip messages
would be exchanged through HTTPS requests and replies, after the TLS Handshake
is completed. This guarantees that an eavesdropping adversary cannot even
distinguish whether a client is gossiping or not. Also, this approach allows to
send longer messages.  The maximum size of an HTTP header is a configuration
parameter, usually set between 4--16~KB. Furthermore, the gossip exchange can
occur for every HTTPS request/reply (with TLS, it happens only during the
handshake).

\myparagraph{Implementation.} To prove the feasibility and efficiency of our
proposals, Protocol~\ref{algo:protocolsc} was implemented and evaluated for both
clients and servers. For log operations, the CT Python API was used. The
server-side component was realized with the Django web framework, and the
client-side consisted of a simple HTTP client that connects to the server,
selects a message, and includes it into an HTTP header dedicated to gossiping.
When an HTTP request is received, the server's middleware checks if it contains
a gossip header and then updates the local storage accordingly. Then, the server
sends a corresponding HTTP reply and embeds the selected message into the gossip
header as well. In turn, the client receives the reply, processes the message,
and updates its storage.

\myparagraph{Performance.} To evaluate the performance of this implementation,
several tests were conducted on a commodity machine (Intel i5-3380M, 2.90 GHz
with 16 GB of RAM) running Ubuntu 14.04, with a 100 Mbps Internet connection.
For the log, one of Google's CT log server was used, and all operations were
executed 100 times. The latency incurred by downloading an STH, a consistency
proof, or an audit proof was on average $162$ ms. The verification of STH
validity, consistency between two STHs, and SCT audit proof took on average
68.87 ms, 0.17 ms, and 0.28 ms, respectively. The most important computational
overhead is caused by the log's signature verification, but it will not be
noticed by users, given that these operations are non-blocking.

\section{Conclusion}
\label{sec:conclusions}
Detecting and disseminating the misbehavior of log servers is the missing aspect
of the current log-based PKI architectures. In this paper, we presented the
first gossip protocols for Certificate Transparency that aim at detecting
several types of attacks by log servers. The concepts that we developed might be
adapted to other public-log approaches~\cite{CIRT, AKI, ARPKI, PoliCert}. Our
proposals do not require any overlay network or dedicated infrastructure, and
can be incrementally deployed.  We evaluated our schemes using real traffic
traces to show that the protocol in which both consistency proofs and STHs are
gossiped is the most promising approach for detecting inconsistencies with a
small overhead. Our research showed that it is possible to implement a gossip
protocol that would greatly improve security with a  small deployment effort and
without sacrificing performance or privacy. Thus, gossip protocols will play an
important role in the upcoming deployment of Certificate Transparency.

\section*{Acknowledgments}
\label{sec:acknowledgments}
This work was made possible by a gift from Google. We thank SWITCH and
Brian Trammell for providing us with the network traces used in the simulation.
We thank Emilia Kasper for her feedback during the initial stage of this work.
We are also grateful to the anonymous reviewers for their valuable comments.

\bibliographystyle{IEEEtran}
\bibliography{ref}

\begin{thebibliography}{10}\vspace{3mm}
\providecommand{\url}[1]{#1}
\csname url@samestyle\endcsname
\providecommand{\newblock}{\relax}
\providecommand{\bibinfo}[2]{#2}
\providecommand{\BIBentrySTDinterwordspacing}{\spaceskip=0pt\relax}
\providecommand{\BIBentryALTinterwordstretchfactor}{4}
\providecommand{\BIBentryALTinterwordspacing}{\spaceskip=\fontdimen2\font plus
\BIBentryALTinterwordstretchfactor\fontdimen3\font minus
  \fontdimen4\font\relax}
\providecommand{\BIBforeignlanguage}[2]{{\expandafter\ifx\csname l@#1\endcsname\relax
\typeout{** WARNING: IEEEtran.bst: No hyphenation pattern has been}\typeout{** loaded for the language `#1'. Using the pattern for}\typeout{** the default language instead.}\else
\language=\csname l@#1\endcsname
\fi
#2}}
\providecommand{\BIBdecl}{\relax}
\BIBdecl

\bibitem{clark:sok}
J.~Clark and P.~C. van Oorschot, ``{SoK}: {SSL} and {HTTPS}: Revisiting past
  challenges and evaluating certificate trust model enhancements,'' in
  \emph{IEEE Symposium on Security and Privacy}, 2013.

\bibitem{comodo-attacks}
P.~Roberts, ``Phony {SSL} certificates issued for {Google}, {Yahoo}, {Skype},
  others,''
  \url{http://threatpost.com/phony-ssl-certificates-issued-google-yahoo-skype-others-032311/},
  2011.

\bibitem{eckersley2010observatory}
P.~Eckersley and J.~Burns, ``An observatory for the {SSL}iverse,'' 2010.

\bibitem{certified-lies}
C.~Soghoian and S.~Stamm, ``\BIBforeignlanguage{English}{Certified lies:
  Detecting and defeating government interception attacks against {SSL} (short
  paper)},'' 2012.

\bibitem{turkey_mitm}
A.~Langley, ``Enhancing digital certificate security,''
  \url{http://googleonlinesecurity.blogspot.co.uk/2013/01/enhancing-digital-certificate-security.html},
  2013.

\bibitem{france_mitm}
------, ``Further improving digital certificate security,''
  \url{http://googleonlinesecurity.blogspot.com.au/2013/12/further-improving-digital-certificate.html},
  2013.

\bibitem{Mazieres:2002:BSF:571825.571840}
D.~Mazi\`{e}res and D.~Shasha, ``Building secure file systems out of byzantine
  storage,'' in \emph{Proceedings of the Twenty-first Annual Symposium on
  Principles of Distributed Computing}, 2002.

\bibitem{Voulgaris:2003:RSP:2155716.2155725}
S.~Voulgaris, M.~Jelasity, and M.~van Steen, ``A robust and scalable
  peer-to-peer gossiping protocol,'' in \emph{Proceedings of the Second
  International Conference on Agents and Peer-to-Peer Computing}, 2004.

\bibitem{1176982}
A.~Ganesh, A.-M. Kermarrec, and L.~Massoulie, ``Peer-to-peer membership
  management for gossip-based protocols,'' 2003.

\bibitem{crosby2009efficient}
S.~A. Crosby and D.~S. Wallach, ``Efficient data structures for tamper-evident
  logging.'' in \emph{USENIX Security Symposium}, 2009.

\bibitem{rfc6962}
B.~Laurie, A.~Langley, and E.~Kasper, ``{Certificate Transparency},'' RFC 6962
  (Experimental), Internet Engineering Task Force, 2013.

\bibitem{ct-ev}
B.~Laurie, ``Improving the security of {EV} certificates,''
  \url{http://www.certificate-transparency.org/ev-ct-plan}, 2014.

\bibitem{population2014}
{U.S. Census Bureau}, ``Countries and areas ranked by population 2014.''

\bibitem{internet-usage2012}
{International Telecommunication Union (Geneva)}, ``Percentage of individuals
  using the internet 2000--2012.''

\bibitem{cameron2013regression}
A.~Cameron and P.~Trivedi, \emph{Regression Analysis of Count Data}, ser.
  Econometric Society Monographs.\hskip 1em plus 0.5em minus 0.4em\relax
  Cambridge University Press, 2013.

\bibitem{CABForum}
{Certificate Authority and Browser (CA/B) Forum}, ``Baseline requirements for
  the issuance and management of publicly-trusted certificates,'' 2013.

\bibitem{Laurie:2014:CT:2668152.2668154}
B.~Laurie, ``Certificate transparency,'' \emph{Queue}, 2014.

\bibitem{overclocking_ssl}
A.~Langley, ``{Overclocking SSL},''
  \url{https://www.imperialviolet.org/2010/06/25/overclocking-ssl.html}, 2010.

\bibitem{CIRT}
M.~D. Ryan, ``Enhanced certificate transparency and end-to-end encrypted
  mail,'' in \emph{Proceedings of the Network and Distributed System Security
  (NDSS) Symposium}, 2014.

\bibitem{AKI}
T.~H.-J. Kim, L.-S. Huang, A.~Perrig, C.~Jackson, and V.~Gligor, ``{Accountable
  Key Infrastructure (AKI): A Proposal for a Public-Key Validation
  Infrastructure},'' in \emph{Proceedings of the International World Wide Web
  Conference (WWW)}, 2013.

\bibitem{ARPKI}
D.~Basin, C.~Cremers, T.~H.-J. Kim, A.~Perrig, R.~Sasse, and P.~Szalachowski,
  ``{ARPKI: Attack Resilient Public-key Infrastructure},'' in \emph{Proceedings
  of the ACM Conference on Computer and Communications Security (CCS)}, 2014.

\bibitem{PoliCert}
P.~Szalachowski, S.~Matsumoto, and A.~Perrig, ``{PoliCert: Secure and Flexible
  TLS Certificate Management},'' in \emph{Proceedings of the ACM Conference on
  Computer and Communications Security (CCS)}, 2014.

\end{thebibliography}

\end{document}